\documentclass[twocolumn,prl,floatfix,showpacs]{revtex4}
\usepackage{graphicx}

\catcode`@=11%
\def\@dotsep{2}%
\catcode`@=12%

\begin{document}
\title{Density of States Monte Carlo Method for Simulation of Fluids}
\author{Qiliang Yan}
\author{Roland Faller}
\author{Juan J. de Pablo}
\email{depablo@engr.wisc.edu}
 \affiliation{Department of Chemical Engineering\\
  University of Wisconsin-Madison\\
  Madison, WI 53706}
\date{\today}

\begin{abstract}
A Monte Carlo method based on a density-of-states sampling is
proposed for study of arbitrary statistical mechanical ensembles
in a continuum. A random walk in the two-dimensional space of
particle number and energy is used to estimate the density of
states of the system; this density of states is continuously
updated as the random walk visits individual states. The validity
and usefulness of the method are demonstrated by applying it to
the simulation of a Lennard-Jones fluid. Results for its
thermodynamic properties, including the vapor-liquid phase
coexistence curve, are shown to be in good agreement with
high-accuracy literature data.
\end{abstract}
 \maketitle

\section{Introduction}
The free energy landscapes of complex systems, such as proteins or
glasses, are characterized by the existence of deep, local free
energy minima. These minima pose significant obstacles for
molecular simulations; once a system is trapped in a minimum,
conventional algorithms are unable to explore configurations
pertaining to other, relevant regions of phase space. Several
Monte Carlo methodologies have been developed in the last decade
to circumvent the sampling problem. Examples include expanded
ensembles\cite{expanded1, expanded2}, multicanonical
algorithms\cite{multicanonical}, and parallel tempering formalisms
\cite{parallel-tempering}. All of these techniques have improved
considerably our ability to simulate the equilibrium properties
and structure of complex materials, including polymers, proteins
in solution, or organic glasses.

Most of these techniques have relied on Metropolis' et al.
original prescription\cite{metropolis}, in which trial
configurations of a system are accepted or rejected according to
probability distributions pertaining to conventional statistical
mechanical ensembles (or minor alterations thereof). A notable
exception is provided by multicanonical methods, in which trial
configurations are accepted according to multicanonical weights
constructed in such a way as to ``flaten out'' high energy
barriers between distinct configurations. Unfortunately, the
required multicanonical weights must be determined through a
tedious and highly computationally demanding iterative process. In
multicanonical methods as well as in more conventional Monte Carlo
techniques for molecular simulation, the condition of detailed
balance is satisfied by construction.

Recently, however, Wang and Landau\cite{dos-1} proposed a
promising approach for lattice systems which eliminates some of
the shortcomings of the original multicanonical prescription. The
key to that approach is to accept trial configurations of the
system according to a \emph{running} estimate of the density of
states; by design, their formalism does not strictly enforce
detailed balance. A random walk in energy space is used to evenly
visit each energy level. The density of states of an energy level
is modified by an arbitrary convergence factor when that energy
level is visited. By controlling that factor in a systematic
manner, these authors were able to generate the density of states
of an Ising lattice system to high accuracy in a self-consistent
way. We refer to this method as Density-Of-States (DOS) Monte
Carlo.

For simulations on a lattice, DOS Monte Carlo promises to offer
significant advantages over previously available techniques. It is
therefore of interest to explore whether analogous ideas can be
used in a continuum, in the context of a realistic fluid. Several
challenges must be overcome: first, random moves on a lattice
system can only give rise to a small set of discrete energy
changes, thereby simplifying considerably the nature of the
simulations. Random displacements in a continuum result in
unpredictable energy changes, and it is unclear whether DOS Monte
Carlo can be implemented at all. Second, and perhaps more
important, DOS Monte Carlo generates an estimate of the density of
states to within a multiplicative constant. Such a constant
depends on the particular length of a simulation, on system size
(volume), and on density (number of particles). On a lattice, it
is common practice to study systems of constant composition.
Furthermore, the system is generally assumed to be incompressible.
Real fluids, however, are not incompressible, and their study over
any reasonable range of density would require that the absolute
density of states (and its multiplicative constant) be known. A
naive extension of the lattice DOS method to a real fluid would
therefore require that multiple simulations be conducted at
different densities, and that the resulting densities of states
corresponding to each calculation be matched with each other in
some fashion to estimate each multiplicative constant. This
procedure would be prone to uncertainties related to the manner in
which different thermodynamic states are combined, it would be
time consuming, and it would not offer significant advantages over
thermodynamic integration or histogram
reweighting\cite{histogram-reweighting}.

In this work we propose a DOS formulation which addresses these
issues. Its implementation is demonstrated in the context of a
Lennard-Jones fluid, for which high-accuracy thermodynamic data
area available. It is shown that the proposed simulation technique
is able to generate the partition function of the system over a
wide range of energy \emph{and} density, which comprises an
infinitely dilute gas and a dense liquid. This is achieved within
a single simulation, at a small fraction of the computational
demands of currently available simulation techniques.

\section{Density of States Monte Carlo}

On a lattice, a DOS Monte Carlo method generates a random walk in
energy space by flipping spins in a random manner. A trial flip is
accepted according to criteria which eventually result in a flat
distribution of energy. Generating a flat energy distribution
requires that microscopic states having internal energy $E$ be
visited with a probability inversely proportional to the density
of states, $g(E)$. Accepting trial spin flips with probability
\begin{equation}\label{goe1}
p(E_1 \to E_2) = \min\left[1, \frac{g(E_1)}{g(E_2)}\right] ,
\end{equation}
(where $E_1$ and $E_2$ are the energies before and after the spin
is flipped), would therefore result in a flat energy distribution.

Unfortunately, the density of states $g(E)$ is not known \'a
priori. The central idea in DOS Monte Carlo is to construct the
density of states on the fly. At the beginning of a simulation, a
constant density of states is assumed for all energy levels. As
the simulation proceeds, the \emph{instantaneous} values of $g(E)$
are used to accept trial flips. Every time that an energy level
$E$ is visited, the density of states for that level is modified
according to $g(E) \to g(E) * f$, where $f > 1$ is an arbitrary
convergence factor. An energy histogram $H(E)$ is accumulated
during the course of the simulation. When that histogram becomes
sufficiently flat, the convergence factor is reduced to a finer
value, \emph{e.g.} $f_{i+1} = \sqrt{f_i}$; the histogram $H(E)$ is
reset to zero, and the simulation is continued. This process is
repeated until the convergence factor $f$ becomes arbitrarily
small (say, less than $\exp(10^{-8})$).

In the many-body fluid formulation proposed here, the number of
particles or molecules in the system is allowed to fluctuate. We
consider a phase space characterized by internal energy $E$ and
number of particles $N$. To sample relevant regions of phase space
efficiently, our strategy is to have each pair of $(N, E)$ points
be visited uniformly. In other words, a microscopic state having
$N$ particles and energy level $E$ should be visited with
probability $1/g(N, E)$, where $g(N,E)$ represents the
two-dimensional degeneracy of $(N,E)$ states.

For simplicity, only two types of trial moves are considered. The
first type of move consists of simple trial displacements, in
which the coordinates of a randomly chosen particle are altered by
a small random amount. A trial move is then accepted or rejected
according to
\begin{equation}
p(E_1,N \to E_2,N) = \min\left[1,
\frac{g(E_1,N)}{g(E_2,N)}\right].
\end{equation}
The second type of move consists of trial insertions or
destruction of particles. For a trial insertion, a particle is
introduced into the system at a random position. This move is
accepted with probability
\begin{equation}
\label{eq:insert}
p(N \to N+1) = \min\left[1,
\frac{V}{(N+1)\Lambda^3}\times\frac{g(N, E^{\mathrm{old}})}{g(N+1,
E^{\mathrm{new}})}\right],
\end{equation}
where $V$ is the volume of the system, and $\Lambda$ is the de
Broglie thermal wave length. In Equation~(\ref{eq:insert}),
$E^{\mathrm{old}}$ and $E^{\mathrm{new}}$ represent the energies
of the system before and after a particle is introduced,
respectively, and $g(N,E^{\mathrm{old}})$ and $g(N+1,
E^{\mathrm{new}})$ are the corresponding densities of states. For
a trial destruction, a randomly chosen particle is removed from
the system, and the move is accepted with probability:
\begin{equation}
p(N \to N-1) = \min\left[1, \frac{N \Lambda^3}{V}\times\frac{g(N,
E^{\mathrm{old}})}{g(N-1, E^{\mathrm{new}})}\right].
\end{equation}

It is of interest to remark that traditional, Metropolis-type
simulation techniques penalize trial insertions or destructions of
particles by a factor proportional to the exponential of the
energy, $ \exp(-\beta \Delta E)$, where $\Delta E$ represents the
change in energy created by the trial move, and where
$\beta=1/k_\mathrm B T$ ($T$ is the temperature). If $\Delta E$
exceeds a few $k_\mathrm B T$, the trial move is generally
rejected. That factor is absent from the algorithm proposed in
this work, thereby facilitating considerably the creation or
destruction of molecules and the sampling of high-density
configurations.

A running estimate of the two-dimensional density of states is
continuously updated as the simulation proceeds. When a
configuration having $N$ particles and energy $E$ is visited, the
current value of $g(N,E)$ is multiplied by a convergence factor
$f$. A two-dimensional histogram of number of particles and energy
$H(N,E)$ is constructed. When that histogram is deemed to be
sufficiently flat, the histogram is discarded and the simulation
is continued, this time with a smaller convergence factor.

Having generated a density of states according to the procedure
outlined above, a partition function for the system can be
constructed at any given temperature and chemical potential
according to
\begin{equation}\label{eq:grand partition}
\Xi(T, \mu)=\sum_N \sum_E g(N,E)e^{-\beta E + N\beta\mu} .
\end{equation}
The partition function determined from Equation~\ref{eq:grand
partition} is only known to within an arbitrary constant
multiplier. Our simulation, however, comprises the special case
$N=0$; the density of states for that case is known to be unity,
thereby providing a means to the determine the absolute value of
$\Xi$ for all other states. Thermodynamic properties of interest
can subsequently be determined from knowledge of $\Xi(T,\mu)$. For
example, the thermodynamic pressure and the thermodynamic internal
energy can be calculated according to
\begin{equation}\label{pressure}
p(T, \mu) = \frac{k_\mathrm B T}{V} \log\Xi(T, \mu)
\end{equation}
and
\begin{equation}
U(T, \mu) = \frac{\displaystyle\sum_N\sum_E E g(N, E) e^{-\beta E
+ N\beta\mu}} {\displaystyle\sum_N \sum_E g(N,E)e^{-\beta E +
N\beta\mu}} .
\end{equation}

The compressibility of the fluid can also be inferred from
$g(N,E)$ by considering fluctuations in the number of particles
according to
\begin{equation}
\kappa(T, \mu) = -\frac{1}{V}\left(\frac{\partial V}{\partial
p}\right)_{N,T} = \frac{V}{k_\mathrm B T}\frac{\langle N^2 \rangle
- \langle N \rangle^2}{\langle N\rangle^2} , \label{eqn:kappa}
\end{equation}
where
\begin{equation}
\langle N\rangle = \frac{\displaystyle\sum_N\sum_E N g(N,E)
e^{-\beta E + N\beta\mu }}{\displaystyle\sum_N\sum_E g(N,E)
e^{-\beta E + N\beta\mu}},
\end{equation}
and
\begin{equation}
\langle N^2\rangle = \frac{\displaystyle\sum_N\sum_E N^2 g(N,E)
e^{-\beta E + N\beta\mu }}{\displaystyle\sum_N\sum_E g(N,E)
e^{-\beta E + N\beta\mu}}.
\end{equation}

In many applications, it is of particular interest to determine
the precise location of first or second order thermodynamic phase
transitions. In the particular case of a simple fluid, to
calculate the liquid-vapor binodal curve, a simple two-state
construction can be used. At a temperature well below the critical
point, the equilibrium density distribution exhibits two distinct
peaks; a threshold number of particles $N_0$ can be designated,
such that all $N \leq N_0$ states can be regarded as pertaining to
a ``vapor'' branch, and all $N > N_0$ states as belonging to a
``liquid'' branch. The pressures corresponding to these two
branches can be calculated according to
\begin{eqnarray}
p^V(T, \mu) = \frac{k_\mathrm B T}{V} \log\sum_{N \leq N_0}\sum_E
g(N, E) e^{-\beta E + N\beta\mu} \label{pe1}\\
p^L(T, \mu) = \frac{k_\mathrm B T}{V} \log\sum_{N > N_0}\sum_E
g(N, E) e^{-\beta E + N\beta\mu} \label{pe2}.
\end{eqnarray}

For any given temperature, a phase coexistence point can be found
by carefully tuning the value of chemical potential $\mu_0$ in
such a way as to satisfy the condition $p^V(T,\mu_0) =
p^L(T,\mu_0)$. The coexistence densities of the vapor and liquid
phases can then be calculated as
\begin{eqnarray}
\rho^V(T) = \frac{\displaystyle\sum_{N \leq N_0}\sum_E N g(N,E)
e^{-\beta E + N\beta\mu_0}}{V\displaystyle\sum_{N \leq N_0}\sum_E
g(N,E) e^{-\beta E + N\beta\mu_0}} \\
\rho^L(T) = \frac{\displaystyle\sum_{N > N_0}\sum_E N g(N,E)
e^{-\beta E + N\beta\mu_0}}{V\displaystyle\sum_{N > N_0}\sum_E
g(N,E) e^{-\beta E + N\beta\mu_0}}.
\end{eqnarray}

Note that the equilibrium pressure obtained from
Equations~\ref{pe1} and \ref{pe2} differs from that obtained from
Equation~\ref{pressure} by a term $-k_\mathrm B T \log 2 / V$,
which arises from finite-size effects. For systems having a large
enough volume, however, this difference is negligible.

One particular problem that must be addressed in the method
proposed above is that the relevant range of energy is strongly
dependent on the number of particles in the system. The energy
levels accessible to a two-particle system are different than
those accessible to a 100-particle system. To determine energy
ranges for different system sizes, two short preliminary DOS
simulations are run, one at the lowest temperature, and the other
at the highest temperature. The goal of these simulations is to
achieve a flat distribution of $N$, as opposed to trying to make
the $(N, E)$ histogram flat; the energy distribution is dictated
by a conventional Boltzmann weight. A flat $N$ distribution would
be obtained if each microscopic state with $N$ particles was
visited with probability $1/Q(N)$, where $Q(N)$ is the canonical
partition function (of a system having $N$ particles).

The scheme followed in these preliminary runs is similar to that
adopted above: a table of partition functions is constructed, with
entries for each value of $N$. At the beginning of a simulation,
the partition functions are set to unity for all entries of $N$.
Two types of moves are used: particle displacements, accepted
according to conventional Metropolis criteria, and particle
insertion or destruction moves, accepted with probability
\begin{equation}
p(N \to N+1) = \min\left[1,
\frac{V}{(N+1)\Lambda^3}\times\frac{Q(N)}{Q(N+1)}\times\exp(-\beta\Delta
E) \right],
\end{equation}
and
\begin{equation}
p(N \to N-1) = \min\left[1,
\frac{N\Lambda^3}{V}\times\frac{Q(N)}{Q(N-1)}\times\exp(-\beta\Delta
E) \right],
\end{equation}
where $\Delta E$ is the energy change associated with the trial
particle insertion or destruction. Upon each visit to a
microscopic state, the corresponding $Q(N)$ is updated by
multiplying it by an arbitrary convergence factor $f$. The minimum
and maximum values of energy corresponding to each number of
particles are tracked during the simulation; the minimum energy at
the lowest temperature and the maximum energy of the highest
temperature define the relevant energy range to be used in the
two-dimensional $g(N,E)$ DOS production simulations.

Since the purpose of the preliminary runs is to determine the
energy range, it is not necessary to generate a perfectly flat
histogram. The only requirement is that each number of particles
be visited with enough frequency. It is important to emphasize,
however, that the scheme described above for $Q(N)$ is very useful
on its own right, for example for calculation of the chemical
potential of a fluid. If this DOS $Q(N)$ simulation is run until
convergence is achieved (i.e. until the $N$ histograms are flat
and the convergence factor $f$ is small enough), the result is the
free energy of the system as a function of $N$. This information
is particularly useful in expanded ensemble
simulations\cite{expanded1, expanded2}, where the weights
associated with individual expanded states (which are closely
related to free energies) must be determined before a production
simulation. In the scheme described here, there is no need to
determine those weights, as they would be determined directly
through the course of the simulation.

\section{Application to Lennard-Jones Fluids}

The remainder of this letter describes results from the
application of the DOS method outlined above to the particular
case of a truncated Lennard-Jones fluid. This model exhibits most
of the main features that one expects to find in most realistic
fluids, and offers the advantage that high-accuracy simulation
data are available for its thermodynamic properties. The potential
energy of interaction between two particles is of the form
\begin{equation}
U(r) = \left\{
\begin{array}{l@{\quad\quad\quad}l}
 0&  r \geq r_\mathrm c\\
 4\epsilon\left[\displaystyle\left(\frac{\sigma}{r}\right)^{12} -
 \left(\frac{\sigma}{r}\right)^6\right]& r < r_\mathrm c \;,
\end{array} \right.
\end{equation}
where $r$ is the distance between the particles, and $r_\mathrm c$
is the cutoff distance. To compare our results to those reported
in the literature, we use $r_\mathrm c = 2.5\sigma$. The box
length of the system is set to $L = 5\sigma$.

Before a simulation is conducted, a relevant range of interest for
the number of particles and the energy must be specified. In this
work, that range is set to be between 0 to 110 for the number of
particles, which covers the density range $\rho^* = 0$--$0.88$.
The range of energy comprises $E/\epsilon=-690$ at one end of the
spectrum, and $E/\epsilon=10$ at the other end; this range
corresponds to temperatures between $0.5<T^*<1.5$.

In contrast to a spin lattice, the potential energy of a
Lennard-Jones fluid is continuous; a discretization of the energy
must therefore be introduced. In this work, an energy bin size of
$\epsilon$ is used to construct the density of states and the
required histograms. Approximately $5\times10^7$ Monte Carlo steps
were used to generate the complete density of states.

\begin{figure}[tbp]
\includegraphics[width = \columnwidth]{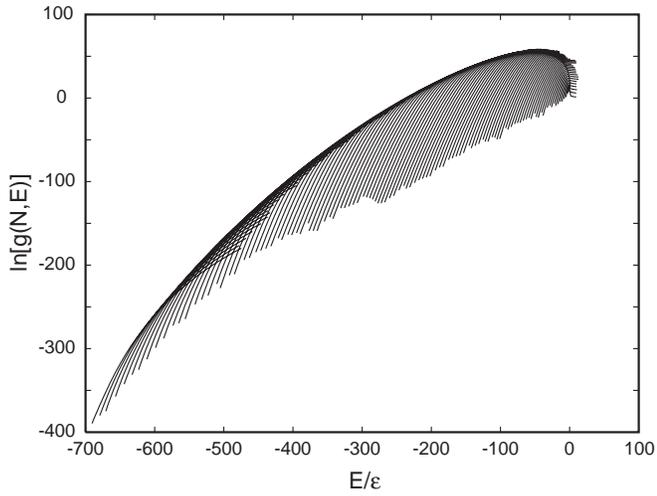}
\caption{\label{g_of_e} Two-dimensional density of states of the
truncated Lennard-Jones fluid. Different lines correspond to a
different number of particles. The number of particles increases
monotonically from right to left.}
\end{figure}

Figure~\ref{g_of_e} shows the density of states as a function of
energy and number of particles. Note that for most $(N,E)$ pairs,
the corresponding density of states is less than unity. This
apparently paradoxical feature is actually due to the fact that
the de Broglie thermal wave length is set to unity in our
simulations.

\begin{figure}
\includegraphics[width = \columnwidth]{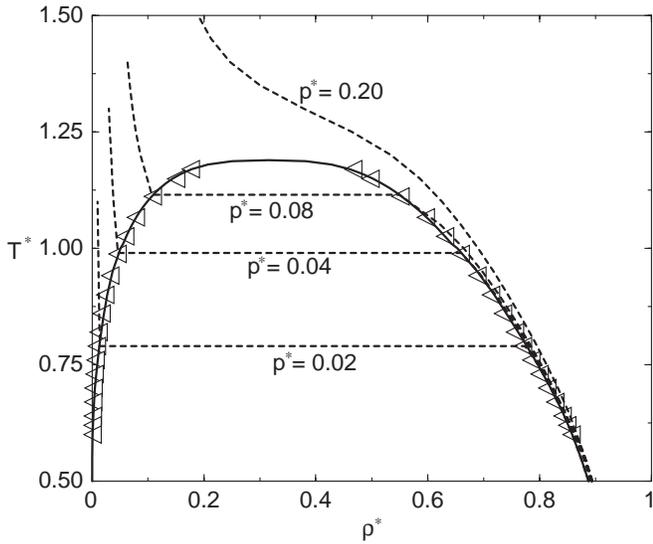}
\caption{\label{phase} Phase diagram of the truncated
Lennard-Jones fluid. The solid line shows the results of this
work; the triangles depict literature data for the same system
\cite{hyper-parallel-1}. The dashed lines are isobars calculated
from the density of states.}
\end{figure}

\begin{figure}
\includegraphics[width = \columnwidth]{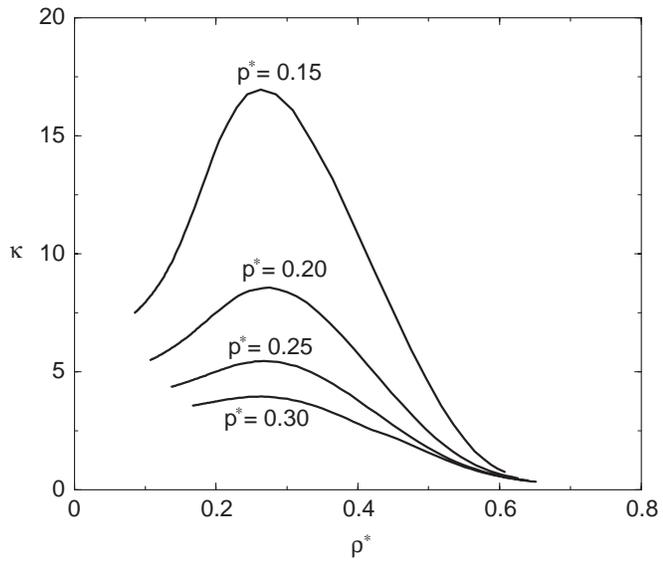}
\caption{\label{kappa} Compressibility of the truncated
Lennard-Jones fluid as a function of density, along various
supercritical isobars.}
\end{figure}

Figure~\ref{phase} shows the phase diagram of the truncated
Lennard-Jones fluid. The solid line was calculated from the
density of states determined in this work; the triangles show
literature results for the same system \cite{hyper-parallel-1}.
The agreement between the two sets of data is good. The figure
shows several isobars, also calculated from the density of states.
These isobars are in excellent agreement with pressure
calculations from conventional canonical-ensemble simulations
(results not shown). The isothermal compressibility of the system,
calculated according to Equation~\ref{eqn:kappa}, is shown in
Figure~\ref{kappa} as a function of density along several isobars.
These results are also consistent with those of conventional
simulations. A distinct peak can be observed in the
compressibility near $\rho=0.3$. As expected, the peak becomes
more pronounced as the pressure approaches the critical pressure
of the system (and fluctuations become more prominent).

\section{Conclusions}

A density of states Monte Carlo method has been presented for
simulation of the thermodynamic properties of realistic fluids
over wide ranges of density and energy. This method permits
calculation of virtually all of the thermodynamic properties of a
system, including its phase behavior, as a function of density or
temperature; all of this information is generated from a
\emph{single} simulation. Furthermore, by virtue of the way in
which trial configurations are generated, states of the system
that would otherwise be difficult to sample (\emph{e.g.} low
temperatures or high densities) can be studied efficiently and
with remarkable accuracy. An additional benefit of the method is
that prior knowledge of the behavior of the fluid (such as
approximate location of phase boundaries) is not required; a
``blind'' simulation is able to sample relevant configurations of
the system with little, if any user-supplied guidance.

For the particular case of a truncated Lennard-Jones fluid, it has
been shown that the procedure proposed in this work generates
thermodynamic information in quantitative agreement with high
accuracy literature data for the same fluid. The coexistence
curve, vapor pressure, and isothermal compressibility are all in
agreement with results of conventional simulation techniques. For
this fluid, the entire range of temperature and density considered
in this work can be generated in several hours of computer time.

While in this work the DOS Monte Carlo method has only been
applied to a simple fluid, our preliminary results for polymeric
systems, glasses, and model proteins indicate that the method
offers considerable promise for simulation of truly complex
fluids.

\begin{acknowledgments}
\end{acknowledgments}

\bibliography{dos}

\newpage
\listoffigures
\newpage

\end{document}